\documentclass{elsart}

\usepackage{amssymb}

\begin{document}

\begin{frontmatter}



\title{Quadratic supersymmetric transformations of the Dirac Green functions}


\author{Ekaterina Pozdeeva}

\address{Department of Quantum Field Theory,
  Tomsk State University, 36 Lenin Avenue, Tomsk, 634050, Russia}

\begin{abstract}
We consider the quadratic supersymmetric aspect of the Darboux
transformation for the Green functions of the one-dimensional Dirac
equation with a generalized form of the potential. We obtain the
relation between the initial and the transformed Green functions on
the whole real line. We also construct the formula for the
unabridged trace of the difference of the transformed and the
initial Green functions of the  boundary problem on the whole real
line.  We present an example illustrated our developments.
\end{abstract}

\begin{keyword}
Dirac equation; Green function; Darboux transformation.

\PACS  02.30.Ik\sep03.65.Ge\sep03.65.Pm
\end{keyword}
\footnotetext[0]{\textit{Email adress:} ekatpozdeeva@mail.ru
  (E. Pozdeeva).}
\renewcommand{\thempfootnote}{\arabic{mpfootnote}}
\footnotetext[1]{supported in part by the "Dynasty" Fund  and Moscow
International Center of Fundamental Physics.}
\end{frontmatter}

\section{Introduction}

The Darboux transformation (DT) \cite{Darbu,Rosu,Song2,Axel1}
provides a very powerful method for constructing  supersymmetric
quantum mechanical systems \cite{Fernand1,maydan1,Gomez,SUZKO}  and
for finding new solvable potentials both for the Schr\"o\-din\-ger
equation \cite{BagrovSamsonov,Sukumar1985,Samsonov1} and the Dirac
equation \cite{Pozdeeva,Eurjphys,BAGROV}.

It is well known that  supersymmetric quantum  mechanics \cite{Witt}
is basically equivalent to DT and the factorization properties of
the Schr\"o\-din\-ger equation
\cite{Gomez,BagrovSamsonov,Sukumar1985}. The DT of the
one-dimensional stationary Dirac equation is equivalent to the
quadratic supersymmetry (QS) and the factorization properties of the
Dirac equation \cite{Eurjphys,BAGROV,annphys2003v305p151}.

Despite of the growing number of papers in this aria the author are
aware of only several papers
 \cite{Sukumar,Samsonov,Pupasov,Sampropagator1,Guenther,PozdeevaGreen} devoted to the supersymmetric
 transformations at the level of the  Green functions.  In \cite{Sukumar} was
obtained  the integral relation between the Green
 functions for two supersymmetric partner Hamiltonians of the one-dimensional
 Schr\"o\-din\-ger equation  with the discrete spectrum.
 In \cite{Samsonov,Pupasov} the integral relation between
the Green functions corresponding two supersymmetric partner
Hamiltonians of the one-dimensional Schr\"o\-din\-ger equation  is
generated to the case of continuous spectrum. The exact the
Schr\"o\-din\-ger propagators
 for the supersymmetric partners are studied \cite{Pupasov,Sampropagator1,Guenther}. In \cite{PozdeevaGreen} the results analogous to \cite{Sukumar} was
obtained for the Dirac equation. In the present paper, we construct
the QS transformations (QST) of the Green function for the whole
real line of the one-dimensional stationary Dirac equation with a
generalized form of the potential and generalize the results of
\cite{PozdeevaGreen} to the case of continuous spectrum.

 The plan of
the paper is as follows. In Section 2 we study the QST of the Dirac
equation. In Section 3 we consider the Green function of the
one-dimensional stationary Dirac equation with a generalized form of
the potential for the whole real line. In Section 4 we construct the
QST of the Green function. We also obtain the relation between the
transformed and the initial Green functions. In Section 5 we get
formulas for the unabridged trace of the difference of the modified
and the initial Green functions. In Section 6 we illustrate our
developments using the free particle examples considered on the
whole real line.
Consequently, in Section 7 we briefly discuss the results.
\section{The QST of the Dirac equation}
In this section we will consider the QST of the one-dimensional
Dirac equation.
We will first look like a particular case of the QST (the matrix
A=I) \cite{annphys2003v305p151}:
\begin{eqnarray} \label{999}Q=\left(\begin{array}{cc}\ 0
& 0\\L & 0\end{array}\right),\qquad  Q^+=\left(\begin{array}{cc}\ 0
& L^+\\0 & 0
\end{array}\right),\qquad
\label{32i} Q^2=(Q^+)^2=0,
\end{eqnarray}
\begin{eqnarray}
\label{31} \{Q,Q^+\}\equiv QQ^++Q^+Q=(H-\lambda_1I)(H-\lambda_2I),
\quad\label{H} H \equiv \left(\begin{array}{cc}\ h_0 & 0\\0 &
h_1\end{array}\right).
\end{eqnarray}

The QST was established on the factorization properties
\cite{BAGROV,annphys2003v305p151}:
\begin{eqnarray}
\label{13} L^+L=(h_0-\lambda_1I)(h_0-\lambda_2I), \qquad
LL^+=(h_1-\lambda_1I)(h_1-\lambda_2I),
\end{eqnarray}
where
\begin{eqnarray}
L=\partial_x-u_xu^{-1} \label{L}, h_0=i\sigma_2+V_0, &
&h_1=i\sigma_2+V_1, V_1=V_0+[i\sigma_2,u_xu^{-1}],
 \end{eqnarray}
the matrix $u$ is the solution of the initial matrix Dirac equation.

Next, let us consider the general case: the matrix $A\ne I$. Let
introduce the matrix with the properties
\begin{eqnarray}
A^{+}=A^{-1},\label{A1}\qquad i\sigma_2A=Ai\sigma_2
\end{eqnarray}
and the operator $\Gamma=AL$.

Consider the operator multiplications $\Gamma^{+}\Gamma$,
$\Gamma\Gamma^{+}$:
\begin{eqnarray}
\Gamma^{+}\Gamma=(h_0-\lambda_1I)(h_0-\lambda_2I),& &
\Gamma\Gamma^{+}=A(h_1-\lambda_1I)(h_1-\lambda_2I)A^{+}\label{Gam1}
\end{eqnarray}
Taking into account (\ref{A1}), we see that the expression
(\ref{Gam1}) is as follows:
\begin{eqnarray}
\Gamma\Gamma^{+}&=&A(h_1-\lambda_1I)A^{-1}A(h_1-\lambda_2I)A^{-1}\label{Gam11},
\end{eqnarray}
where
\begin{eqnarray}
& &A(h_1-\lambda_1I)A^{-1}=Ah_1A^{-1}-\lambda_1I, \\
& &Ah_1A^{-1}=
i\sigma_2+AV_0A^{-1}+A[i\sigma_2,u_xu^{-1}]A^{-1}-i\sigma_2A_xA^{-1}.
\end{eqnarray}

Let us denote
\begin{eqnarray}
&&H_0=i\sigma_2+W_0\label{HH0}, H_1=i\sigma_2+W_1 \label{HH1},\\
\label{W1}&&W_1=AW_0A^{-1}+A[i\sigma_2,u_xu^{-1}]A^{-1}-i\sigma_2A_xA^{-1},
\end{eqnarray}
\begin{eqnarray}
 \mathcal{H} \equiv
\left(\begin{array}{cc}\ H_0 & 0\\0 & H_1\end{array}\right),\qquad
 \label{OOO999}\Omega=\left(\begin{array}{cc}\ 0
& 0\\ \Gamma & 0\end{array}\right),\qquad
\Omega^+=\left(\begin{array}{cc}\ 0 & \Gamma^+\\0 & 0
\end{array}\right).
\end{eqnarray}
Now it is easy to obtained that the properties for the operators
$\Gamma$, $\Gamma^{+}$, $\mathcal{H}$, $\Omega$, $\Omega^+$ are
analogous to  the properties  for operators $L$, $L^{+}$, $H$, $Q$,
$Q^+$:\begin{eqnarray}
&&\Gamma^{+}\Gamma=(H_0-\lambda_1I)(H_0-\lambda_2I),\label{Gam0}\qquad
\Gamma\Gamma^{+}= (H_1-\lambda_1I)(H_1-\lambda_2I),\label{gam1}\\
\label{Omega}&& \Omega^2=(\Omega^+)^2=0, \label{O31}\quad
\{\Omega,\Omega^+\}\equiv
\Omega\Omega^++\Omega^+\Omega=(\mathcal{H}-\lambda_1I)(\mathcal{H}-\lambda_2I).
\end{eqnarray}

The QS transformed potential (\ref{W1}) early  was introduced in
\cite{annphys2003v305p151} as the Darboux transformed potential of
the one-dimensional Dirac equation.
\section{Green function of the one-dimensional Dirac equation}
In this section we consider some properties of the Green functions
of the one-dimensional Dirac equation with the generalized form of
the potential.  We will reconcentrate mostly the case on  the whole
real line ($a=-\infty$, $b=\infty$).

Let us consider the one-dimensional Dirac equation
\begin{eqnarray}
\label{equation1} H_0\psi=E\psi,& &x\in (a,b).
\end{eqnarray}
The Hamiltonian $H_0$ include  the generalized form of the potential
 \begin{eqnarray}\label{3}
   W_0(x)=\omega(x)I+(m+S(x))\sigma_3+q(x)\sigma_1,
\end{eqnarray}
where $\omega(x)=\phi_{el}(x)$, $S(x)=\phi_{sc}(x)$,
$q(x)=k_j/x+\phi_{am}$, $k_j$ are the eigenvalues of the spin-orbit
operator \cite{thaler}, $m$ is the mass of a particle, $\sigma_1$,
$\sigma_3$ are usual Pauli matrices. The generalized potential is
the self-adjoint potential.

There are exist two representation of the Green function of the
one-dimensional Dirac equation. The first representation is obtained
with the help of two real solutions $\psi$, $\varphi$.
 The Green function in this case is
\begin{eqnarray} \label{G}
G_0(x,y,E)=\frac{1}{W(E)}\left\{\begin{array}{cc}\
\psi(x,E)  \varphi ^T(y,E), & y\leq x,\\
 \varphi(x,E)\psi^T (y,E), &
x<y,\end{array}\right.
\end{eqnarray}
where $W(E)=W\{\psi(x,E),\varphi(x,E)\}=const$
\cite{Eurjphys,annphys2003v305p151} is the Wronskian of the two
functions $\psi(x)$ and $\varphi(x)$. Early similar view of the
Green functions  was obtained in \cite{levitan} for the Dirac
equation with a particular potential.

If we consider the Green functions on the whole real line, then
suppose that $\psi$, $\varphi$
  satisfy the  zero boundary conditions:
$\varphi(a,E)=0$, $\psi(b,E)=0$.

The matrix (\ref{G}) can be written in the  form:
\begin{eqnarray}
\label{7} G_0(x,y,E)=\frac{\psi(x) \varphi ^T(y)\Theta(x-y)+
\varphi(x)\psi^T (y)\Theta(y-x)}{W\{\psi(x),\varphi(x)\}},
\end{eqnarray}
where $\Theta(x-y)$, $\Theta(y-x)$ are the Heaviside  step
functions.

We suppose that discrete spectrum
eigenfunctions $\{\psi_n\}$,  together with continuous spectrum
eigenfunctions $\{\psi_k\}$, $E=k^2$, is complete in the Hilbert
space
\begin{eqnarray}\label{complete}
    \sum_n\psi_n^T(x)\psi_n(y)+\int dk \psi_k^T(x)\psi_k(y)=\delta(x-y).
\end{eqnarray}
The spectral representation of the Green function may be found in
terms of the complete set as follows:
\begin{eqnarray}\label{spectralrepresentation}
    G_0(x,y,E)=\sum_n\frac{\psi_n(x)\psi^T_n(y)}{E_n-E}+\int\frac{\psi_k(x)\psi^T_k(y)}{k^2-E}dk.
\end{eqnarray}

For the spectral problem on the whole real axis the integrals over
$k$ run from minus infinity to infinity.
\section{The QST of the Green functions}
In this Section using spectral the representation we obtain a
relation between the Green functions of the initial and the
transformed problems. With the help of the properties of the QST we
check that this relation is correct for usual representation of the
Green functions  of the one-dimensional stationary initial and
transformed Dirac equations.

The functions $\tilde{\psi_n}=\Gamma\psi_n$, $n=1,2,...,$ describe
(unnormalized) bound states  and $\tilde{\psi_E}=\Gamma\psi_E$
correspond to (unnormalized) scattering states of $H_1$. The
normalization constants are easily obtained with the help of the
factorization property (\ref{Gam0}). The functions
\begin{eqnarray}
  \phi_n=\frac{\Gamma \psi_n}{\sqrt{(E_n-\lambda_1)(E_n-\lambda_2)}},
  & &\phi_E=\frac{\Gamma \psi_E}{\sqrt{(E_n-\lambda_1)(E_n-\lambda_2)}}.
\end{eqnarray}
form an orthonormal set.

It is well known \cite{annphys2003v305p151} that there exist kinds
of the QST: (i) the transformation deleting the state levels, (ii)
the  isospectral transformation, (iii) and the transformation
creating new state levels.


In the case (i) the spectral representation of the transformed Green
function is given by:
\begin{eqnarray}\label{Greeni}
    G_1(x,y,E)=\sum_{n_1}\frac{\phi_{n_1}(x)\phi^T_{n_1}(y)}{E_n-E}
    +\int\frac{\phi_{k}(x)\phi^T_{k}(y)}{k^2-E}dk.
\end{eqnarray}
The designation $n_1<n_0$ correspond to numbers of the Hamiltonian's
$H_1$ discrete eigenvalues. In the case (ii) $n_1$ is equal $n_0$.
In the case (iii), when the columns of matrix $(u^{+})^{-1}$ are
square-integrable, the spectral representation of the transformed
Green function is as follows:
\begin{equation}\label{Greeni}
    G_1(x,y,E)=\sum_{n_1}\frac{\phi_{n_1}(x)\phi^T_{n_1}(y)}{E_n-E}+
    \sum_{m}\frac{\phi_{\lambda_m}(x)\phi^T_{\lambda_m}(y)}
    {\lambda_m-E}+\int\frac{\phi_{k}(x)\phi^T_{k}(y)}{k^2-E}dk.
\end{equation}
If the transformation creates one new state level, then $m=1$. If
the transformation creates two new state levels, then $m=2$.

For all free case (i), (ii), (iii) the transformed and initial Green
functions are interrelated:
\begin{eqnarray}
\label{GxyE}
G_1(x,y,E)=\frac{L_xG_0(x,y,E)L^T_y-(\gamma\partial_x+V_1+E-
\lambda_1-\lambda_2)\delta(x-y)}{(E-\lambda_1)(E-\lambda_2)}.
\end{eqnarray}

In the fist representation the transformed Green functions  looks
like as
\begin{eqnarray}
\label{G1} G_1(x,y,E)=\frac{\tilde{\psi}(x) \tilde{\varphi}
^T(y)\Theta(x-y)+ \tilde{\varphi}(x)\tilde{\psi}^T
(y)\Theta(y-x)}{W\{\tilde{\psi}(x),\tilde{\varphi}(x)\}}.
\end{eqnarray}

If we consider the Green functions on the whole real line, then
suppose that $\tilde{\psi}$, $\tilde{\varphi}$
  satisfy the  zero boundary conditions:
$\tilde{\varphi}(a,E)=0$, $\tilde{\psi}(b,E)=0$.

For the solutions to the Dirac equation the following properties
hold:
\begin{eqnarray}
\label{phipsi}&&\psi\varphi^T-\varphi\psi^T=-W\{\varphi,\psi\}\gamma,\\
\label{tildaW}
&&W\{\tilde{\varphi},\tilde{\psi}\}=(E-\lambda_1)(E-\lambda_2)W\{\varphi,\psi\},\\
\label{lambdaIlambda2}
&&(u_xu^{-1}-(u_xu^{-1})^T)\gamma=\lambda_1+\lambda_2.
\end{eqnarray}

Since the differential of the Heaviside step function is the
delta-function, the properties (\ref{phipsi}), (\ref{tildaW}),
(\ref{lambdaIlambda2}) are fulfilled. Substituting (\ref{7}) into
(\ref{GxyE}) we obtain Eq. (\ref{G1}).

\section{Unabridged trace for the
 difference  of the modified and the initial Green functions}
In this Section we consider the  unabridged trace for the Green
function defined as  $tr\int^b_aG(x,y,E)|_{x=y}dy$. It can be
represented in the form:
\begin{equation}\label{115}
     tr\int^b_aG_1(x,y,E)|_{x=y}dy=
\frac{tr\int^b_a\tilde{\psi}\tilde{\varphi}^T|_{x=y} dy}
{W\{\tilde{\psi}\tilde{\varphi}\}}.
\end{equation}

It is known that the action of the transformation operator  and the
conjugate transformation one on the spinors can be written in the
following way \cite{annphys2003v305p151}:
\begin{eqnarray}
  \label{Lpsi1}
   L\psi&=& u\frac{d}{dx}(u^{-1}\psi),\\
  \label{Lpsi1+}
    L^+\tilde{\psi}&=& -(u^{+})^{-1}\frac{d}{dx}(u^{-1}\tilde{\psi}).
\end{eqnarray}
Accounting these properties, we can write (\ref{115}) in the form:
\begin{equation}\label{1153}
      tr\int^b_aG_1(x,y,E)|_{x=y}dy=
tr\int^b_a\frac{d}{dx}(u^{-1}\psi)\tilde{\varphi}^Tu|_{x=y}
 dy\biggl/W\{\tilde{\psi}\tilde{\varphi}\}.
\end{equation}
We integrate (\ref{1153}) by parts and apply the trace property
\begin{equation}\label{1152}
tr{AB}=tr{BA}
\end{equation}
to obtain the relation
\begin{equation}\label{1154}
tr\int^b_aG_1(x,y,E)|_{x=y}dy=
\frac{tr\psi\tilde{\varphi}^T|^b_a+\int^b_a tr\psi(L^\dagger
 L\varphi)^T|_{x=y}dy}{W\{\tilde{\psi}\tilde{\varphi}\}}.
    \end{equation}
Due to (\ref{tildaW}) and the factorization property from
\cite{annphys2003v305p151} we obtain:
\begin{equation}\label{1155}
\int^b_atrG_1(x,y,E)|_{x=y}dy=\frac{tr\psi\tilde{\varphi}^T|^b_a}
{W\{\tilde{\psi}\tilde{\varphi}\}}+\int^b_a trG_0(x,y,E)|_{x=y}dy.
\end{equation}
Similarly, we would like to write
\begin{equation}\label{1157}
\int^b_atrG_1(x,y,E)|_{x=y}dy=\frac{tr\tilde{\psi}\varphi^T|^b_a}
{W\{\tilde{\psi}\tilde{\varphi}\}}+\int^b_a trG_0(x,y,E)|_{x=y}dy.
\end{equation}
  Finally, from (\ref{1155}), (\ref{1157}) the formulae
for the unabridged trace of the
 difference  of the modified and  the initial Green functions are the following:
\begin{eqnarray}
\label{TIG1}
tr\int^b_a(G_1(x,y,E)-G_0(x,y,E))|_{x=y}dy&=&\frac{1}{W\{\tilde{\psi},
\tilde{\varphi}\}}tr(\psi\tilde{\varphi}^T)\mid^b_a,\\
\label{TIG}
tr\int^b_a(G_1(x,y,E)-G_0(x,y,E))|_{x=y}dy&=&\frac{1}{W\{\tilde{\psi},
\tilde{\varphi}\}}tr(\tilde{\psi}\varphi^T)\mid^b_a.
\end{eqnarray}
Here $a=-\infty$, $b=\infty$ are the limits of integration.

\section{Example}
First of all, we would like to note the multiplication property
$\psi(x)\varphi(y)^T$:
\begin{eqnarray}
(\psi(x)\varphi(y)^T)^T=\varphi(y)\psi(x)^T.
\end{eqnarray}
The multiplication $\varphi(x)\psi(y)^T$ is obtained due to the
transformation of the multiplication $\psi(x)\varphi(y)^T$ and the
substitution $x\longleftrightarrow y.$ Analogously,
\begin{eqnarray}\psi(x)\varphi(y)^T=(\varphi(x)\psi(y)^T)^T\:(x\longleftrightarrow
y).\end{eqnarray}

 Now let us  consider the examples illustrating the results obtained
 above.\\
1. We assume the free motion case $V_0=m\sigma_3$ on whole real line
$x\in[-\infty,\infty]$. We chose the solutions of initial solutions
as follows:
\begin{eqnarray}
\psi=\left(
  \begin{array}{c}
    \exp(-ikx) \\
      ik\exp{(-ikx)}/(E+m)\\
  \end{array}
\right),\qquad \varphi=\left(
  \begin{array}{c}
    ik\exp{(ikx)}/({E-m}) \\
     \exp{(ikx)}\\
  \end{array}\right).
\end{eqnarray}
Use these solutions we construct the Green function:
\begin{eqnarray}
G_0(x,y,E)=\frac{1}{2}\left\{\mathcal{A}^{(0)}\Theta(x-y)+\mathcal{B}^{(0)}\Theta(y-x)\right\},\end{eqnarray}
where
 \begin{eqnarray}\mathcal{B}^{(0)}=\left(\begin{array}{cc} -ik/(E-m)&-1\\
 1&\qquad
ik/(E+m)\end{array}\right)\exp{i(kx-ky)},\end{eqnarray}
 \begin{eqnarray}\mathcal{A}^{(0)}=(\mathcal{B}^{(0)})^T(x\longleftrightarrow y),& \qquad &k=\sqrt{E^2-m^2}.\end{eqnarray}

The matrix solution of the initial Dirac equation can be chosen
\cite{annphys2003v305p151} as follows:
\begin{eqnarray}\label{uuu}
    u=\left(%
\begin{array}{cc}
  1 &\qquad {\rho}\sinh{(\rho x)}/(\varepsilon-m)\\
  0 & \qquad \cosh{\rho x} \\
\end{array}%
\right),
\end{eqnarray}
where $\lambda_1=m$, $\lambda_2=\varepsilon<m$,
$\rho=\sqrt{m^2-\varepsilon^2}$. The corresponding potential are
follows
\begin{eqnarray}\label{V111}
    V_1(x)=-\varepsilon\sigma_3+\rho\tanh{(\rho
    x)}\sigma_3,
\end{eqnarray}
the function $-\varepsilon$ is  playing  a part of  mass.

Use the QST transformation we construct the solutions of the Dirac
equation with potential (\ref{V111}):

\begin{eqnarray}
\tilde{\psi}=\frac{ik}{E+m}\left(
  \begin{array}{c}
    \varepsilon-E\\
      -ik-\rho\tanh{(\rho x)}\\
  \end{array}
\right)\exp(-ikx) ,\end{eqnarray}

\begin{eqnarray}
\tilde{\varphi}= \left(
  \begin{array}{c}
      \varepsilon-E \\
      ik-\rho\tanh{(\rho x)}\\
  \end{array}\right)\exp{(ikx)}.
\end{eqnarray}
Now we construct the QST transformed Green function corresponding
the potential (\ref{V111})
\begin{eqnarray}
G_1(x,y,E)=\frac{\mathcal{A}^{(1)}\Theta(x-y)+\mathcal{B}^{(1)}\Theta(y-x)}{2k(E-m)(E-\varepsilon)},
\end{eqnarray}
where
\begin{eqnarray}
\mathcal{B}^{(1)}=\left(\begin{array}{cc}\mathcal{ B}^{(1)}_{11}& \mathcal{B}^{(1)}_{12}\\
\mathcal{B}^{(1)}_{21} &
\mathcal{B}^{(1)}_{22}\end{array}\right)\frac{ik}{E+m}\exp{i(kx-ky)},
\end{eqnarray}
\begin{eqnarray} \mathcal{A}^{(1)}&=&(\mathcal{B}^{(1)})^T(x\longleftrightarrow
y),\\
\mathcal{B}^{(1)}_{11}&=&(\varepsilon-E),\\
 \mathcal{B}^{(1)}_{21}&=&(ik-\rho\tanh{(\rho x)})(\varepsilon-E)\\
 \mathcal{B}^{(1)}_{12}&=&(\varepsilon-E)(-ik-\rho\tanh{(\rho y)})\\
\mathcal{B}^{(1)}_{22}&=& -(ik-\rho\tanh{(\rho
x))}(ik+\rho\tanh{(\rho y})).
\end{eqnarray}
Apply the formula of the unabridged trace from difference
transformed and initial  Green functions (\ref{TIG1}), we compute
the following trace:
\begin{eqnarray}
tr\int^b_a(G_1(x,y,E)-G_0(x,y,E))|_{x=y}dy=-\frac{i\rho}{k(E-\varepsilon)}.
\end{eqnarray}
The expression reflect the addition one energy level to the spectrum
of the modified Hamiltonian.

\section{Conclusion}

 In this paper we studied the QST of the Green function and the relations between the
transformed Green functions and the initial Green function  of the
one-dimensional Dirac equation for the case the whole real line.
 We also construct  the formulas for the unabridged trace of the difference of the transformed  and the
initial Green functions of the one-dimensional Dirac equation for
the case are the whole real line. The main results of the paper are
the relation between the initial and the transformed Green functions
(\ref{GxyE}) and the formulae for an unabridged trace (\ref{TIG1}),
(\ref{TIG}) in the whole real line. We case of a half-line is more
complicated. We believe that this case will be investigate in a
separate publication.

\section*{Acknowledgments}

The author is grateful to  Dr. D. Antonov for helpful comments of
the given work. I would like also the express my thanks to Prof. B.
G. Bagrov for useful discussions. This work was supported in part by
the ``Dynasty'' Fund and Moscow International Center of Fundamental
Physics.

\end{document}